# The self-referring DNA and protein: A remark on physical and geometrical aspects

Tsvi Tlusty

September 4, 2015

*Simons Center for Systems Biology, Institute for Advanced Study, Princeton, NJ 08540, USA*

*Center for Soft and Living Matter, Institute for Basic Science (IBS), Ulsan 689-798, Korea*

*Department of Physics, Ulsan National Institute of Science and Technology, Ulsan 689-798, Korea*

## Abstract
All known life forms are based upon a hierarchy of interwoven feedback loops, operating over a cascade of space, time and energy scales. Among the most basic loops are those connecting DNA and proteins. For example, in genetic networks, DNA genes are expressed as proteins, which may bind near the same genes and thereby control their own expression. In this molecular type of self-reference, information is mapped from the DNA sequence to the protein and back to DNA. There is a variety of dynamic DNA-protein self-reference loops, and the purpose of this remark is to discuss certain geometrical and physical aspects related to the back-and-forth mapping between DNA and proteins. The discussion raises basic questions regarding the nature of DNA and proteins as self-referring matter, which are examined in a simple toy model.

## Introduction

To self-reproduce, living matter must refer to itself. This was lucidly depicted in von Neumann's treatment of self-reproducing automata [1], where he described a cellular automaton whose grid cells are identical finite-state machines. The cells may take several states or 'phenotypes', and their spatial organization forms a machine with means of storing, reading, transmitting, and processing digital information. In each 'generation', the machine refers to a 'tape' of such cells, which stores a blueprint of actions to be performed by the machine in order to reproduce a complete copy of itself.

In von Neumann's automaton, both roles of information storage and its logical processing are performed by the same kind of cells via their action as finite state machines. In the biological realm, this is similar to the ability of certain RNA molecules to self-replicate [2]. This led some to suggest a primordial RNA world occupied by evolving ensembles of autocatalytic and self-reproducing RNAs. However, all existing or known to have existed organisms share a dichotomy between the blueprint drawn in DNA molecules, and the machine-like proteins that execute the blueprint. RNAs still carry out essential roles of catalysis (ribosome), information transfer (mRNA) and control (small RNAs), which mediate between the biochemical worlds of DNA and protein.

The schism into two styles of biochemistry allows DNA and proteins to excel in their specialized tasks. DNAs are linear, easy-to-manipulate hetero-polymers, whose inertness secures the *digital* information written in the four letter alphabet of the nucleic bases[1]. Proteins, on the other hand, are folded chains of amino acids that evolved to perform catalysis, signal transduction and other highly specific functions. The possible configurations of proteins may be specified by the 3D position of each amino acid, and their biochemical functions are quantified by parameters such as affinities and catalysis rates. These biochemical and configurational degrees of freedom reside in high-dimensional *analogue* (continuous) spaces.

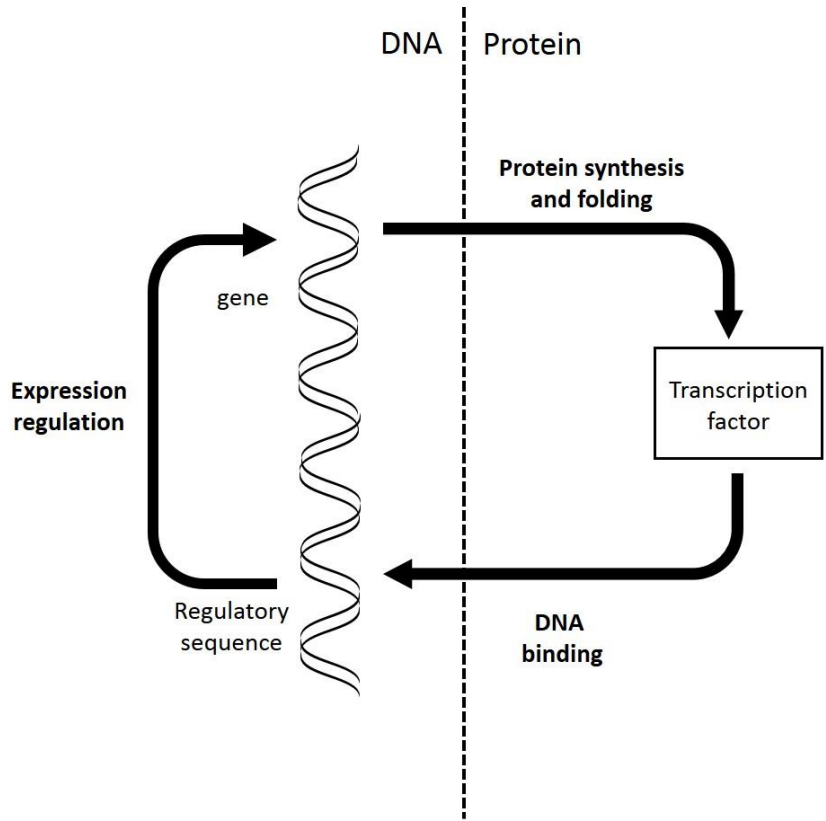

Figure 1: **A gene autoregulation loop crosses twice the DNA-protein boundary.** A DNA gene is written transcribed and translated to make a transcription factor protein (square). In the case of autoregulation, the transcription factor binds the regulatory sequence adjacent to its own gene, thereby controlling its own expression. The dashed line denotes the boundary between the DNA and protein worlds. Crossing this boundary requires translation or mapping.

The dual worlds of protein and DNA are interconnected by numerous biochemical pathways, and this back and forth exchange generates feedback loops of various length and complexity. For example, a gene written in DNA is expressed, via transcription and translation, as a protein, which in turn may bind the DNA as a transcription factor, thereby controlling its own expression or that of other proteins in the cell (Figure 1). Such autoregulation loops are prevalent in genetic networks. Negative control stabilizes expression and leads to homeostasis, whereas positive loops may act as switches. The biochemical and evolutionary aspects of DNA-protein loops of various architecture were extensively studied in systems biology (e.g., [3-7]). Here we focus on certain geometrical aspects of the mapping between DNA and proteins and their implication on living systems.

---

[1] The DNA encodes also analogue biophysical properties, such as affinities of the corresponding mRNA to certain proteins and small RNAs, and the interaction with nucleosomes. These properties can be interpreted as secondary codes.

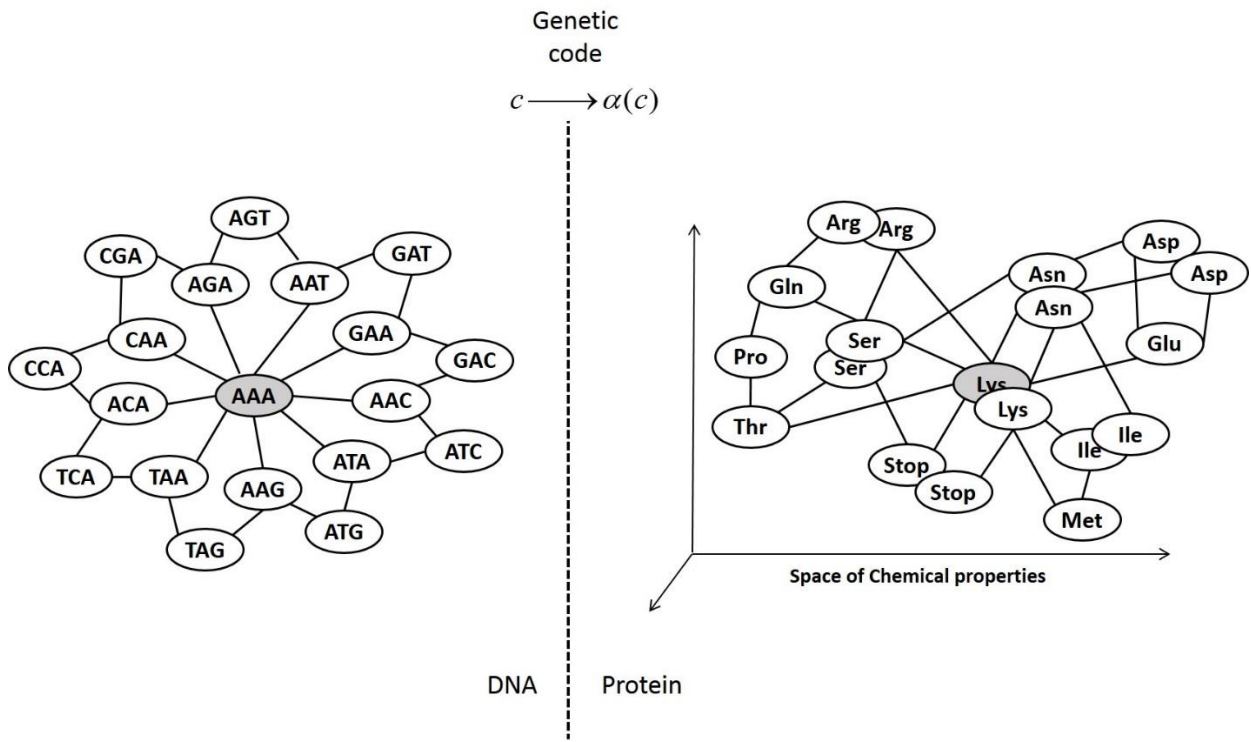

Figure 2: **the map of the genetic code.** Left: A 19 codon sub-graph of the 64 codon graph showing the codon c = AAA with its nine nearest neighbors, and nine of its tqwnty-seven next-nearest neighbors. The whole 64-codon graph is highly interconnected, and cannot be drawn on a plane without many edge intersections. Right: the genetic code maps each codon c to an amino acid α(c), denoted by a three-letter code, or a stop signal. The amino acids reside in a space whose axes correspond relevant chemical properties such as size, polarity, hydrophobicity etc. The map deforms the graph. Neighboring synonymous codons, which are mapped to the same amino acid, reside at the same position and the corresponding connecting edge shrinks, whereas amino acids of different chemical properties are far from each other with long connecting edges.

## The loop and the representation problem

The loop described above (Figure 1) maps back and forth between the dual molecular worlds of DNA and proteins. The elements or 'atoms' of these two worlds are the nucleic bases along the DNA, and the amino acids along the protein chain. On this elementary 'atomic' level, the genetic code maps DNA triplets called codons into distinct amino acids. We previously discussed this map in detail [8, 9], and here we mention briefly its prominent geometric features.

The codon space can be described as a graph whose vertices are the $4^3 = 64$ codons (Figure 2). Two vertices are connected by an edge if the two codons are likely to be confused by the translation machinery. Possible errors include misreading of the RNA codon in the ribosome, mischarging of the tRNA by the wrong amino acid, and point mutations of the DNA. In the resulting structure, called the Hamming graph, each codon is connected by nine edges to those nine other codons that differ by just one letter. In this graph, the length of the shortest path between two codons measures their number of different letters Hamming distance.

The genetic code maps each codon $c$ into an amino acid $α(c)$, which resides in the space of chemical properties. The map $α(c)$ is known to be rather smooth, in the sense that neighboring codons differing by one base tend to be mapped to the same or chemically similar amino acids. Smoothness increases the tolerance of the genetic code to errors in this information channel, for example due to misreading of a codon at the ribosome. The map $α(c)$ embeds the codon graph in the Euclidean space of chemical properties.

The genetic code is one of the most basic examples of representation or mapping in biology. The digital DNA triplets represent the information about the corresponding amino acids in the space of chemical properties. Whenever information is translated or mapped from one type chemical or physiological language to another one, it is being represented in the new language. For example, the retina represents visual information as neural spikes, and signaling molecules represent information regarding the presence of neighboring bacteria in a colony.

On the 'atomic' level of the genetic code, several scenarios and models were suggested to explain the origin of the code and its salient features (e.g., [10, 11] and many others). While many questions remain open, the basic geometry is clear: the genetic code map is basically an embedding or a representation of the Hamming graph into a high dimensional metric space of amino acid chemistry. The smoothness of the code sets an upper bound on the dimension of the chemical property space and thereby on the number of amino acids the code can embed without compromising too much its reliability [8].

## The representation of proteins

Let us now illustrate in slightly more detail the autoregulation loop (Figure 3). On the level of a whole protein, the map from DNA to amino acids is much more elaborate and much less is known about the fundamental properties of this representation. The representation includes a linear stage in which the gene, a sequence of codons $\boldsymbol{c} = (c_1, c_2 ... c_n)$ is translated at the ribosome into a chain of corresponding amino acids $\boldsymbol{α(c)} = (α(c_1), α(c_2) ... α(c_n))$, termed polypeptide. The chain then folds into a 3D configuration of the protein $\boldsymbol{p_c}$, which is formally denoted as $\boldsymbol{p_c} = \boldsymbol{s} \circ \boldsymbol{α(c)}$; the letter $s$ hints for 'structural' degrees of freedom, and $\circ$ denotes composition of functions $\boldsymbol{s} \circ \boldsymbol{α(c)} = \boldsymbol{s(α(c))}$. This stage is highly nonlinear, since the configuration $\boldsymbol{p_c}$ is determined by interactions of the amino acids among themselves and with the surrounding medium, such as hydration.

In a simplified view, $\boldsymbol{p_c}$ can be considered as a vector of the positions of each amino acid (or each atom on a higher resolution) in the protein's native state. Real proteins, however, fluctuate among an ensemble of possible configurations, and intrinsically disordered proteins lack even an average 3D structure [12, 13]. On top of that, it is not clear whether one can, even in principle, deduce a configuration, $\boldsymbol{p_c} = \boldsymbol{s} \circ \boldsymbol{α(c)}$, from the sequence $\boldsymbol{c}$ alone (Anfinsen's dogma [14]). Nevertheless, on the simplified abstract level considered here, $\boldsymbol{p_c}$ can be taken of as a vector whose entries correspond to the structural degrees of freedom, which characterize the protein molecule. In the traditional structural view of proteins, those may be the amino acid positions, while on a more stochastic view one may consider the probability distribution of structures. A convenient representation may be chosen according to the biophysical context. For example, one may expand the deviations from an equilibrium structure as the modes of an elastic network [15].

Describing the detailed structure of a protein requires numerous degrees of freedom, whose number increases with the desired resolution. A dynamical description would require even more degrees of freedom to specify the force fields in the protein, for example the spring constants of an elastic network model. However, it seems that not all those parameters are necessary to understand the function of the protein. In our simple example of an autoregulation loop, the functionality of the protein is characterized by its binding site (in Figure 3, the cluster of dark 'amino acids').

The function of the transcription factor in binding the DNA is governed by the interaction of the DNA with this rather small cluster of amino acids $b_c$. This does not at all imply that the rest of the protein does not matter. On the contrary, the structure and the stability of the binding site depends on its interactions with the 'rest' of the amino acids. Certain binding sites physically interact with distant amino acids, thereby affecting their functional properties. This mechanism known as allostery is central to protein function [16]. The map $f$ formally denotes the process in which certain functional degrees of freedom $b_c$ emerge from the physical interactions of a given protein configuration $s$, $b_c = f \circ s \circ \alpha(c)$. As already mentioned, in general a small set of degrees of freedom are governed by the whole protein. In this sense, the map $f$ is similar to the phenomenon of *renormalization* in physical systems. In broad terms, renormalization may be thought of as a recursive coarse graining process in which one obtains a small number of effective interaction parameters that account for the original, microscopic system parameters [17].

It is noteworthy that, while in physical systems it is clear how to obtain the renormalized degrees of freedom, in the present context of proteins the concept of renormalization is at this point more of a metaphor than a realistic model. The essence of the map $f$ is the drastic *dimensional reduction* in the space of possible configurations. In the following, we present a toy model that aims to capture this essence. While the physical mechanisms which result in dimensional reduction are yet to be elucidated, proteins exhibit properties which are consistent with this concept: The biochemical function of proteins is effectively described by a small number of parameters, for example, many enzymes are characterized by their binding affinity and catalysis rate (the Michaelis-Menten model) [18]. The dynamics that occur during binding and catalysis appear to involve mostly a few low-frequency, large-scale modes [19, 20]. Low modes in the correlations of amino acid substitution during evolution were found to be related to the function and stability of several protein families [21]. It is not clear, however, whether these two types of low modes, evolutionary and dynamic, are related.

## Closing the loop

The interaction of the regulatory sequence $c^r$ with the binding site $b_c$ is formally denoted by the function $\beta$. A simple model for $\beta$ may specify the probability that $c^r$ is bound to a transcription factor $p_c$, while more detailed models add also the binding and unbinding probabilities (a two state Markov model) up to a molecular dynamics simulation. In general, the process of stochastic binding involves diffusion, docking, which depends on the interaction of the binding surfaces, and the induced conformational changes [22, 23]. Yet, in many cases $\beta$ can be described as the sigmoidal binding probability of a two-state model, $\beta = 1/[1 + \exp(\mu - E_B)]$, where $E_B$ is the binding energy and $\mu$ is the chemical potential [24]. While the shape of $\beta$ is non-linear, the binding energy itself can be reasonably approximated as a linear sum of binding energies of each base-pair in the regulatory sequence $E_B(c^r) = \sum_i E(c^r_i)$. In this respect, the binding function $\beta$ is linear in the DNA sequence $c^r$ just as the translation $\alpha$ is linear in the gene sequence $c$. Usually, there is a non-negligible probability that the transcription factor binds to sequences close enough to the 'consensus' sequence $c^r$.

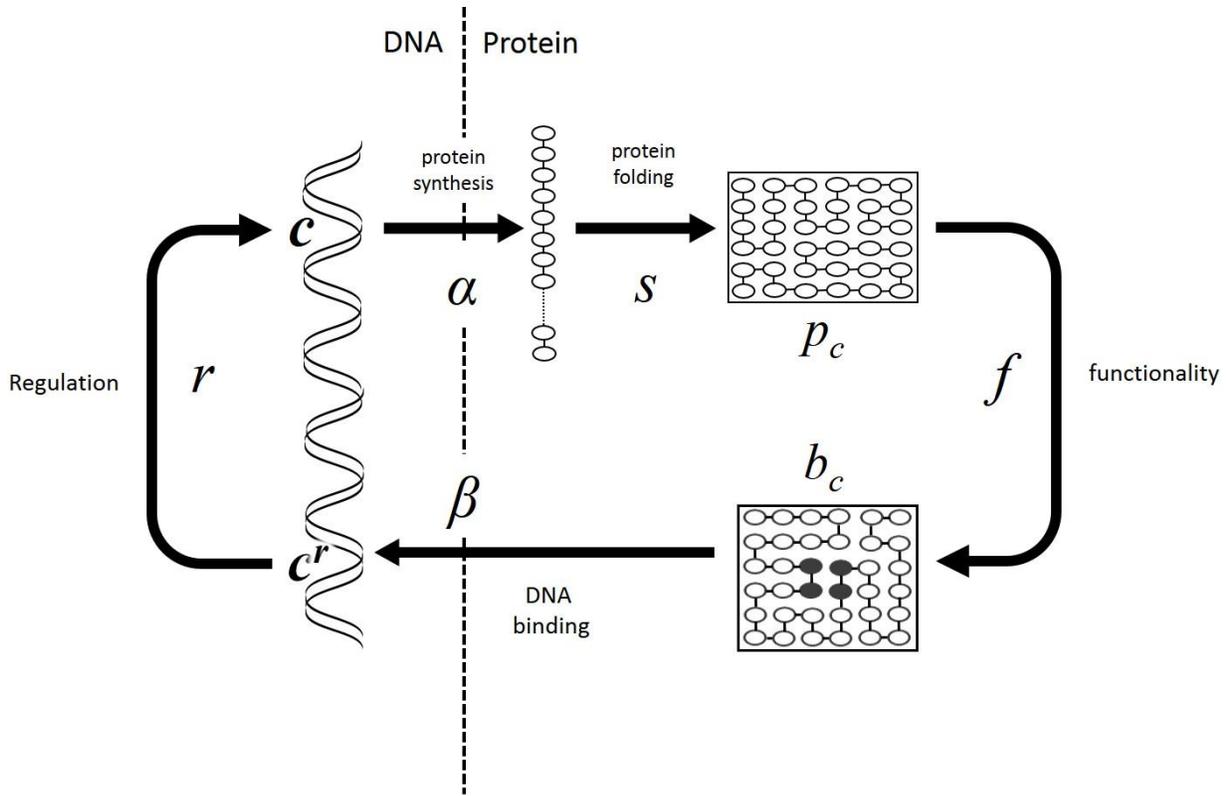

Figure 3: **following the maps along the autoregulation loop.** (i) A gene **c** is a sequence of codons $c = (c_1, c_2 ... c_n)$, which is translated at the ribosome into a chain of corresponding amino acids $\alpha(c) = (\alpha(c_1), \alpha(c_2) ... \alpha(c_n))$, according to the genetic code. The genetic code map $\alpha(c)$ is linear. (ii) Next, the amino acid hetero-polymer is folded into a protein $p_c$. This is formally denoted by a nonlinear map **s**. (iii) The function of the protein $p_c$ as a transcription factor is facilitated by its binding site (dark amino acids). The structure and the chemical properties of the binding site $b_c$ depend on the protein configuration **s**. This map is formally denoted by **f**. (iv) The binding of the protein to the DNA is described by the map **β**, which depends on the binding site $b_c$ and the regulatory sequence $c^r$. (v) The binding or unbinding of the transcription factor regulates the expression of proteins, as described by the regulation map **r**. This stage closes the autoregulation loop. The maps by the genetic code **α** and by the binding **β** are the ones that cross back and forth from DNA to proteins.

The autoregulation loop is closed by the regulatory effect of transcription protein $p_c$ on the expression of its own gene **c**. The regulation is described by the function **r**, which often depends on the local concentration of the transcription factor. The functional dependence exhibits many forms, positive, negative, linear and non-linear, all according to the biological function of the loop, such as stabilization or switching. Formally, the dynamics of the loop is a composition of the functions described above:

$$\frac{d}{dt}[p_c] = r \circ \beta \circ f \circ s \circ \alpha(c, c^r). \tag{1}$$

Equation (1) is deterministic dynamics for the concentration of the transcription factor $[p_c]$, but it could as well be a stochastic equation for protein number. Similar formal description applies also to longer loops, such as a switch made of two proteins controlling the expression of each other (Figure 4):

$$\frac{d}{dt}\left[p_{c_1}\right] = r_1 \circ \beta_2 \circ f_2 \circ s_2 \circ \alpha_2(c_2, c_1^r),$$
$$\frac{d}{dt}\left[p_{c_2}\right] = r_2 \circ \beta_1 \circ f_1 \circ s_1 \circ \alpha_1(c_1, c_2^r). \tag{2}$$

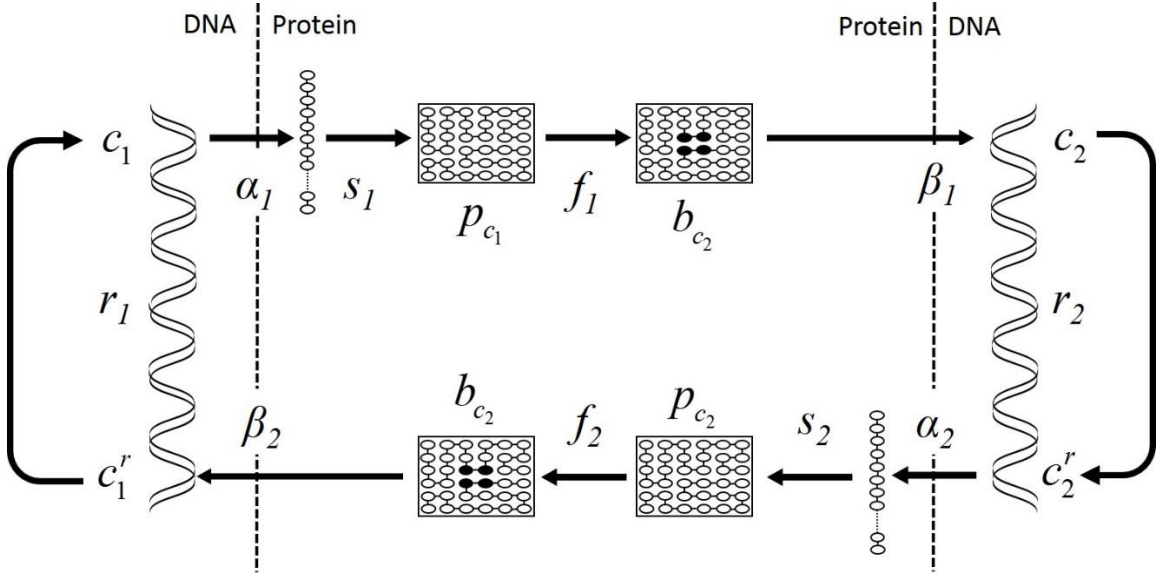

Figure 4: **a two-gene switch.** Two genes, $c_1$ and $c_2$, express proteins, which bind to each other's regulatory sequence. With negative regulation function, $r_1$ and $r_2$, the resulting loop acts as a switch (equation (2)).

When the two regulatory control functions, $r_1$ and $r_2$, are negative, equation (2) becomes a switch.

To conclude, we followed schematically the flow of information from the DNA to the protein and back in the autoregulation loop, where a gene controls its own expression. The pathway composes five maps or functions, $\alpha, s, f, \beta,$ and $r$, which operate on the DNA sequences, the gene $c$, and the regulatory binding site $c^r$. One may mix the maps of several genes to create larger loopy networks, such as that of equation (2). In the following, we discuss the dimensionality along this series of maps, and construct a toy model with similar characteristics.

## Dimensional reduction and expansion – hats and bowties

To retrace the mapping along the loop, let us start again from the gene $c$ (Figure 5). A gene $c$ of length $n$ codons resides in a discrete space of approximately $64^n$ possible configurations. The actual number is somewhat smaller, around $61^n$, as there are 3 stop codons that can appear only as a punctuation signal at the end of the gene. The dimension of the genetic space is proportional to $n$, and its structure can be described as the Cartesian graph product of $3n$ tetrahedra, each corresponding to one base-pair [8, 9].

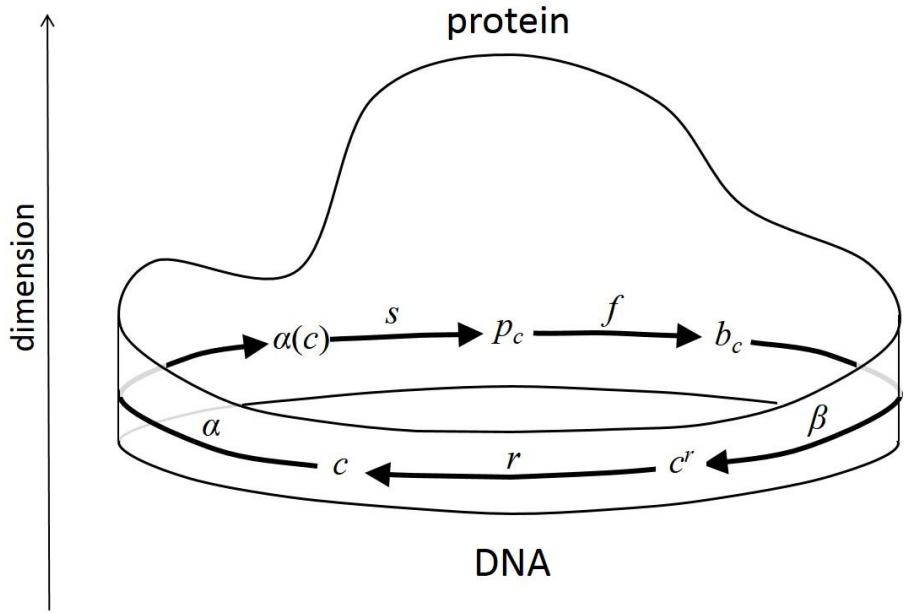

Figure 5: **Tracing the dimensionality along the DNA-protein loop** (see text). The height represents the effective dimension, which expands and shrinks in the protein section of the loop (arrows). The map $\alpha$ slightly reduces the dimension due to the redundancy of the genetic code, as manifested in a small dip at $\alpha(c)$. The map $s$ into the space of structures and dynamical trajectories significantly expands the effective dimension. The following map $f$ filters the functionally relevant information, the structure and biophysical properties of the binding site, thereby drastically reducing the effective dimension. The binding interaction is approximately linear, with an effective dimension that scales like the length of the regulatory sequence.

The genetic code map $\alpha$ slightly reduces this dimensionality due to the redundancy of the code: there are only 20 amino acids which are encoded by 61 codons. Hence, most amino acids are encoded by two or more codons, all synonyms for the same amino acids. The resulting number of possible polypeptides is about $20^n$. However, even synonymous codons yield different mRNAs, which interact differently with the tRNA reservoir and noncoding RNAs in the cell [25]. Therefore, a map which considers also RNA interactions would be of a higher dimension. Unlike the inherently digital space of DNA sequences, the amino acids reside in a space of chemical characteristics, such as polarity, size and hydrophobicity (Figure 2), and the number of relevant characteristic determines the effective dimension. At any rate, both spaces of DNA genes $c$ and amino acid sequences $\alpha(c)$ are linear (in other words, they are product spaces) and their dimension is proportional to $n$.

In principle, one may specify the 3D folding of a protein by listing the coordinates of each amino acids. Moreover, since the amino acids are concatenated into a polypeptide chain, it suffices to list the two torsion angles along the backbone of the polypeptide (Ramachandran plot [26]). However, this would only provide a static structure averaged over an ensemble of protein configurations, and disordered proteins lack even this static average [12, 13]. Proteins are dynamic objects which are strongly coupled to their biochemical surrounding. Therefore, understanding protein function requires much more information regarding the force fields, interaction with the solvent, and the potential conformational changes. Even a coarse grained description in terms of an elastic network, which is valid only for small deformations, requires knowledge of the effective spring constants and the connectivity of the network. The output from the elastic network is a list of $n$ vibrational modes, each of which describes the deviations of the $n$ amino acids from their equilibrium position, and the effective dimension therefore scales like $n^2$.

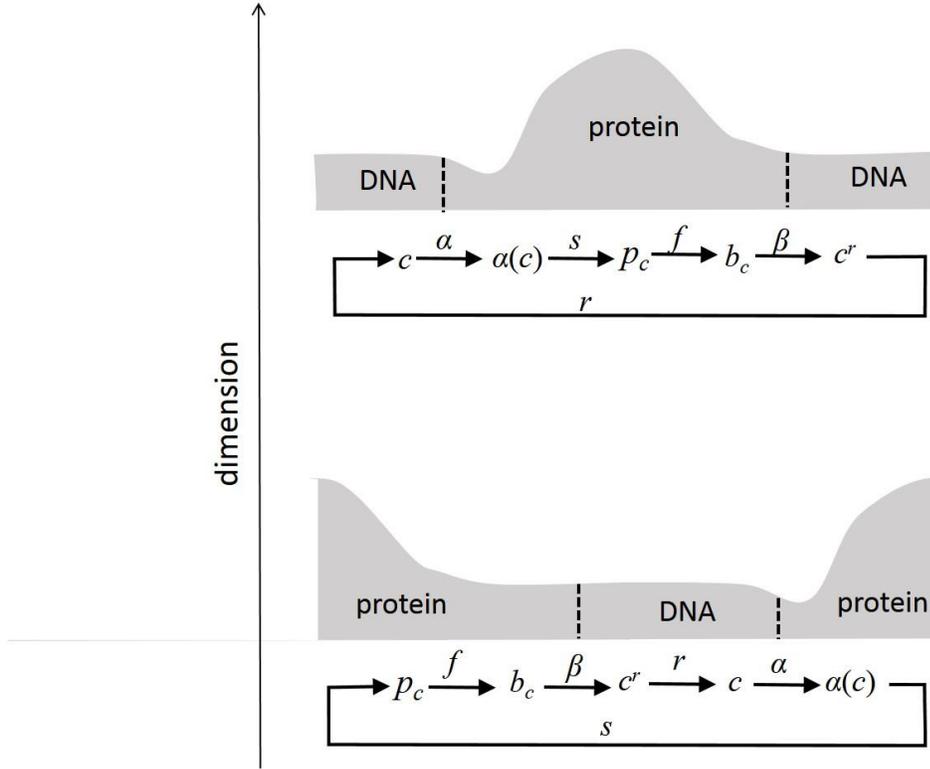

Figure 6: **hats and a bowties.** Depending on where one starts to traverse the autoregulation loop, it may look like a 'hat' (top) or a 'bowtie'. However, such distinction is arbitrary just like the choice of starting point.

Characterization of large conformational changes, which are essential to protein function, requires tracing the trajectories in protein configuration space. Such knowledge is far beyond the standard crystal structure, and is much harder to obtain [27, 28]. This difficulty is reflected in the struggle to advance molecular dynamics modeling of protein beyond the microsecond regime [29-31]. Even an ad-hoc, low resolution description in terms of the amino acid two-body correlation functions has dimension which scales like $n^2$, similar to the scaling of the elastic modes. More elaborate dynamical models, which include trajectories would require a much higher dimension to represent the rich dynamics.

Similar dimensional expansion is observed in the dynamics of spatial networks of Boolean functions or neural networks. While the static description in terms of an interacting spatial network is linear in the number of points, if the range of interaction is limited and thereby the connectivity of each point is bounded. However, their dynamics in state space is much more complicated and requires a super-linear number of parameters. We conclude that the corresponding map $s$ considerably expands the dimension of the space possible states (Figure 5).

The drastic dimensional expansion by the map $s$ is followed by a dimensional reduction of similar magnitude via the map $f$. In the present example, the function of the protein as a transcription factor is governed by its rather small binding site (Figure 3), which can be characterized by relatively small set of parameters, such as the binding affinity, catalysis rate, and the elastic response. In principle however, this small set of numbers depends on distant amino acids, as evident from mutation studies. Another evidence of long-range correlations is the phenomenon of allostery, in which binding of a ligand to one site is modified by binding of another ligand to a distant site (which perhaps may be thought of as a 'viscoelastic transistor'). As mentioned above, the map *f* involves a coarse-graining or renormalization-like process, in which the physical properties of the whole protein are integrated into a small number of effective parameters. There is certain evidence that these effective coordinates correspond to the slow, large-amplitude modes of the protein motion [19, 32, 33].

In the next stage, the binding site of the transcription factor $b_c$ interacts with the regulatory sequence. Often this interaction is reasonably approximated by a linear energy, in which the interaction with each base-pair is independent. At this level, the effective dimension after the binding map *β* is proportional to the length of the sequence $c^r$. The specific binding is a recognition process, which provides information regarding the base-pairs along the regulatory sequence. Following the linear energetics, the information content is also a sum of independent contributions from each base-pair [34].

The last stage involves the genetic regulatory function *r*, which is typically described as a simple relation between the transcription factor concentration and the expression rate of their corresponding gene [3, 4]. The shape of this function is characterized by a few parameters, such as saturation level and nonlinearity (Hill's coefficient). At any rate, this requires only a handful of numbers and the corresponding dimensionality is therefore low.

Dimensional compression and expansion is common to many biological systems. It has been recently suggested that a 'bowtie' structure, that is low-dimensional intermediate stages between inflated input and output spaces, emerge when an evolutionary task can be compressed [35]. The back and forth trajectory between DNA and proteins (Figure 6) may appear more like a 'hat' (or a boa constrictor digesting an elephant) than a bowtie. But this distinction is quite arbitrary in this case, since in principle one could start the loop from the protein to make it appear like a bowtie, although starting at the gene seems more natural. Anyway, the basic feature is a chain of inflationary bubbles of high dimension separated by deflationary low-dimensional bottlenecks. Compression and expansion are typical in other types of information channels and artificial learning systems.

## A toy model

At this point, we can recapitulate the basic geometric aspects of mapping between genes and proteins:

(a) The genetic information for synthesizing a protein is encoded in a *linear* sequence of *n* codons, which are practically independent. The resulting configuration space of all possible sequences is therefore a product space of dimension *n*.
(b) The genetic code preserves this linear scaling of the dimension, with slight reduction due to the redundancy of the code (only 20 amino acids are encoded by 61 non-stop codons).
(c) The dynamics of proteins is governed by the 3D spatial configuration of its amino acids. The short-range interactions among the amino acids yield long-range correlations.
(d) The space of all possible dynamic trajectories scales super-linearly with *n*: specifying all infinitesimal perturbations or two-body correlations requires $n^2$ numbers, and full trajectories would require even higher dimension.
(e) The functionality of a protein is determined by a few effective parameters and a few slow dynamical modes.
(f) The interaction of protein and DNA is roughly linear in the size of the regulatory site.

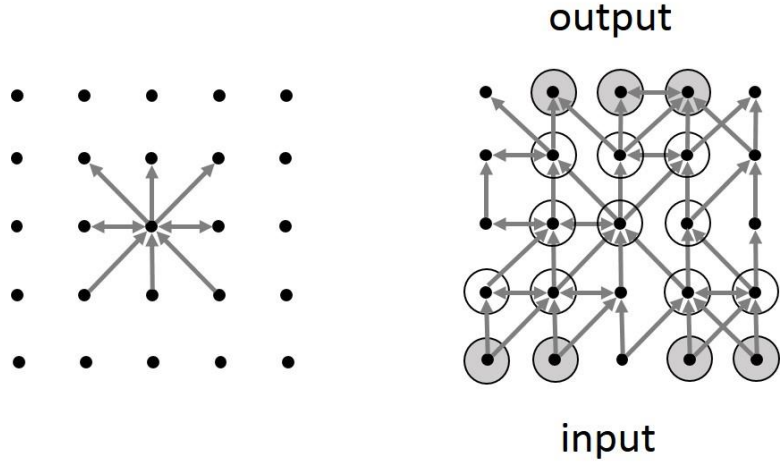

Figure 7: **A toy model.** The 'protein' is made of *n* vertices ('amino acids') ordered is a 2D square lattice with periodic boundaries (cylindrical geometry). **Left**: Each vertex may be connected by directed edges to any of its nearest eight neighbors. Input arrows can come only from five neighbors, the three below and the two on the sides. Similarly, output arrows can point to the three neighbors upwards and to the two on the sides. The network can be represented as an adjacency matrix $A$, in which $A_{ij} = 1$ if an edge points from vertex $j$ to vertex $i$ and $A_{ij} = 0$ otherwise. **Right**: Each vertex $i$ can be either 'on' ($S_i = 1$, denoted by a large clack circle), or 'off' ($S_i = 0$). The states of the vertices are updated according to a 'quorum' rule: if a quorum of at least $k$ vertices that point to $i$ are 'on', then $i$ is also 'on' (equation (3)). The steady-state is shown for $k = 2$. Due to the directed connectivity, the firing front propagates upward from the low 'input' row to the top 'output' row (large gray circles).

In the following, we describe a toy model that mimics these basic features of the DNA-protein system. The toy model is intentionally constructed to bear merely the abstract similarity via features (a)-(f), while one should abandon the illusion that it simulates the physics of real proteins (Figure 7). As a protein-like structure, we consider a network of $n$ vertices (the 'amino acids') ordered in a square lattice of side $\sqrt{n}$ with periodic boundary condition on the left and right sides (each layer is a ring in a cylinder). Each vertex may be connected by directed edges to any of its nearest eight neighbors, and this connectivity determines the spatial architecture of the network. It can be written in terms of an adjacency matrix $A$, where $A_{ij} = 1$ if an edge points from vertex $j$ to vertex $i$ and $A_{ij} = 0$ otherwise.

Each vertex $i$ can be in one of two states $S_i$, 'on' ($S_i = 1$) and 'off' ($S_i = 0$). The dynamics is defined by the following update rule: Count the number of 'on' vertices that point to vertex $i$; if there are $k$ or more 'on' vertices, then $i$ turns on. In terms of the adjacency matrix this rule takes the form:

$$S_i(t+1) = \begin{cases} 1 & : \sum_j A_{ij} S_j(t) \geq k \\ 0 & : \quad \text{otherwise} \end{cases}. \qquad (3)$$

At $k = 1$ the dynamics is that of standard percolation, and the requirement for $k > 1$ active leads to collective dynamics called bootstrap or quorum percolation [36], which was recently applied to living neural networks [37, 38]. The edges can point only in the five upward directions (Figure 7), and information therefore propagates from bottom to top. Hence, the state of the bottom layer defines the 'input' and the top layer is the 'output' of the protein. The transmission of information through the 'protein' is akin to an allosteric effect which couples distant sites in real proteins.

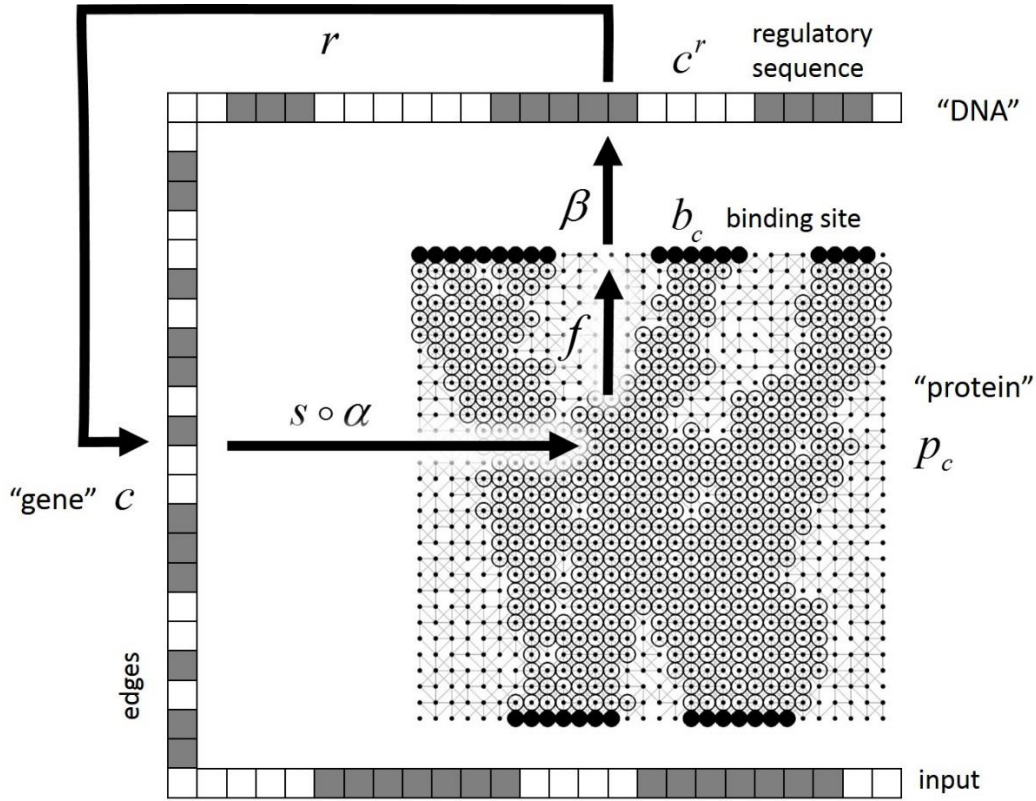

Figure 8: **a toy autoregulation loop.** The 'gene' $c$ encodes in 0's and 1's (light and dark squares) the 'input' (firing pattern of the bottom layer, black circles), and the edge configuration (gray lines in the 'protein'). For each vertex $i$, the gene encodes the five potential inward edges $A_{ij}$. The map from 'gene' to the spatial organization of the 'protein' combines the maps $s$ and $\alpha$. The map $f$ represents the firing dynamics, which determines the binding site $b_c$ ('output') as a function of the structure and the input for $k = 2$. The regulatory sequence $c^r$ is the upper part of the 'DNA'. The binding site $b_c$ interacts with $c^r$ according to the binding map $\beta$. Finally, the regulatory map $r$ controls the expression of the gene as a function of the $b_c$ - $c^r$ interaction, thereby closing the autoregulation loop.

At $t = 0$, vertices in the bottom layer start to fire according to the input pattern. The bootstrap dynamics is monotonic: A vertex that turned 'on' will stay 'on', and the state configuration in the 'protein' $S_i$ will reach a steady-state. The configuration of the top layer at steady state is the 'output' of the protein, and the input-output relation defines the *function* of the network. In the following, we construct a 'DNA-protein' system that exhibits the generic features (a)-(f) of real autoregulation loops.

The DNA-like structure is a linear string of 0's and 1's, composed of a 'gene' $c$ and a regulatory sequence $c^r$ (Figure 8). The 'gene' encodes (i) the 'input' pattern of the 'protein' at its bottom layer, i.e. which of the bottom vertices is 'on', and (ii) the connectivity of the 'protein' network. Since the inward edges are restricted to the five forward and sideways directions (Figure 7), one can encode the whole network in a linear sequence of 0's and 1's, where each vertex $i$ has a quintet of $A_{ij}$ corresponding to the five potential inward edges. The overall length of the 'gene' $c$ is therefore $5n + n^{1/2}$.

In this example, the map from the linear 'gene' $c$ to the architecture of the 2D 'protein' $p_c$ combines the maps $s$ and $\alpha$. One may think of $\alpha$ as a non-degenerate genetic code, which is simply a one-to-one binary representation of the edges and the input. The function of the protein is defined by the pattern of its binding site $b_c$ ('output'), which is the outcome of the collective dynamics $f$ described by equation (3). The map $f$ reduces the effective dimension from that of 'protein' dynamics to $n^{1/2}$, the number of vertices in the 'output' layer. The binding site $b_c$ is indeed a small $n^{-1/2}$ fraction of the vertices, but it is the outcome of the dynamics inside the whole 'protein'. In this sense, the binding site's degrees of freedom 'renormalize' the degrees of freedom in the rest of the protein.

The binding map $\beta$ defines the interaction of the binding site $b_c$ with the regulatory sequence $c^r$. A simple expression for the steady-state binding probability is the sigmoidal curve:

$$\beta = \frac{1}{1 + e^{(\mu - E_B)/T}}, \tag{4}$$

with a linear binding energy, $E_B = \sum_i \delta(c^r_i, b_c^i)$, which counts the number of matches between $b_c$ and $c^r$, and a chemical potential $\mu$. The regulatory map $r$ uses the information from the $\beta$ interaction to modify the expression of the 'gene'. This closes the autoregulation loop.

T he performance of the loop can be tweaked by varying its DNA representation: the input pattern and the network connectivity $A_{ij}$, encoded in the 'gene' $c$, which together determine the output at the binding site $b_c$, or the regulatory sequence $c^r$, which affects the binding and thereby the expression of the gene through the regulatory map $r$. In principle, the maps themselves can also be tweaked by altering, for example, the genetic code $\alpha$, and the physical interaction of folding $s$, or binding $\beta$. However, such changes are expected to be much slower due to their global impact on many DNA-protein pathways. We demonstrate such 'evolutionary' dynamics, where the network $A_{ij}$ adapts towards an optimal input-output relation via random breaking and forming of edges in the 'protein' (Figure 9). A more detailed and quantitative will be reported elsewhere.

## Discussion - What is a protein?

The essence of the protein is in the non-linear mapping from the digital information of the gene to the analog realm of the protein: the spatial configuration of the amino acids, their collective interaction, and the resulting biochemical function, binding to the DNA. Within the abstract simplified view presented here, the DNA-protein mapping was divided into two consecutive maps, the 'structural' map $s$ which corresponds to the folding, the ensemble of protein conformations, and the resulting amino acid interaction, among themselves and with the surrounding solvent. The following map $f$ represents the emergence of functionally relevant dynamical modes in the protein, which is similar in spirit to the standard analysis of physical systems in terms of low-energy, large-amplitude modes.[39].

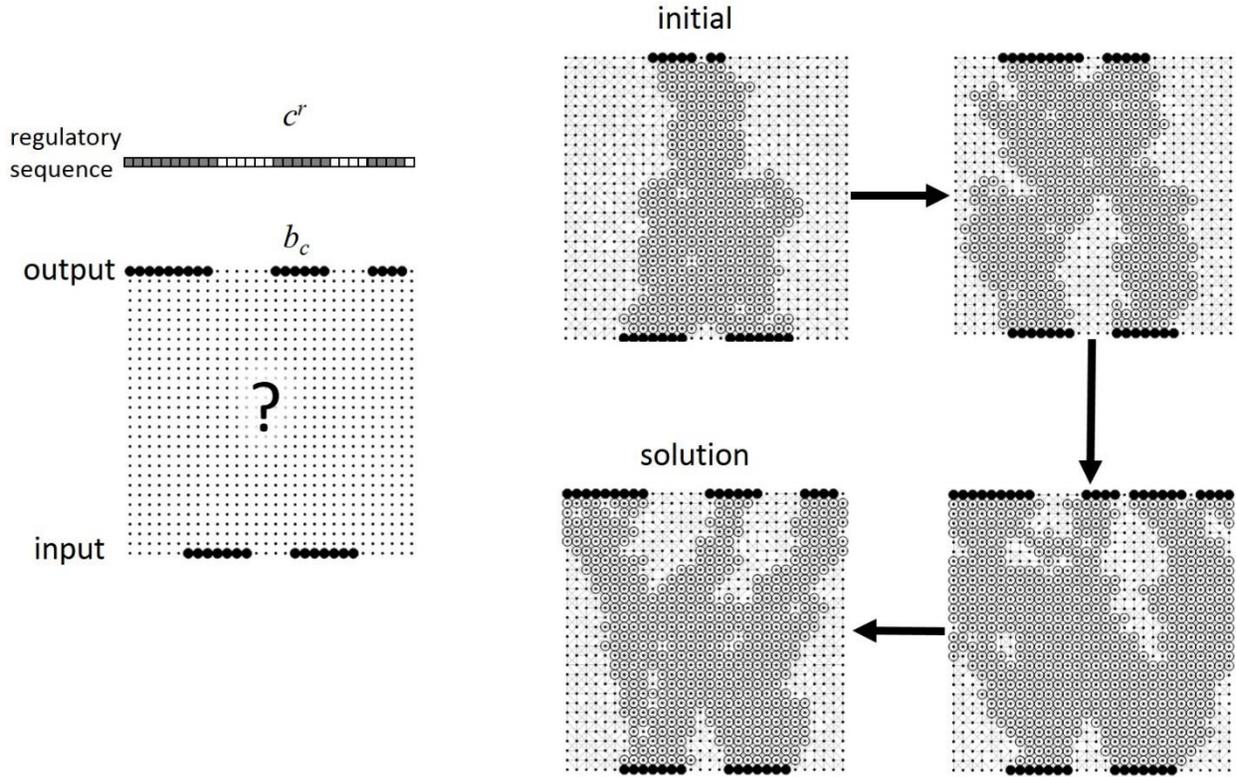

Figure 9: **evolution of a toy protein.** Simulation of evolutionary adaptation of the 'protein' in the autoregulation loop. Left: The performance of the protein in the loop is governed by the interaction of its binding site $b_c$ with the regulatory sequence $c^r$. Optimal performance is achieved when the $b_c$ - $c^r$ matching is maximal. Right: By mutating the network $A_{ij}$, one edge at a time, the 'protein' evolves from an initial random configuration towards a high performance configuration, via intermediate configurations (shown are two configurations out of a few thousands). In the simulation, edges are mutated randomly, keeping mutations that increase the target function of maximal binding energy $E_B$. The solution configuration obeys the optimal input-output configuration.

The division into 'structural' and 'functional' maps may appear analogous to the traditional view of structural biology that the function of the protein relies on its three-dimensional structure. For example, enzymes will bind to specific substrates if their shapes match according to the 'key and lock' principle. However, already the 'induced fit' model [22] demonstrated the critical role of large conformational changes in the function of many proteins [23]. Furthermore, many proteins are known to be functional despite their inherent disorder and the lack of 'structure' in the traditional manner [40, 41]. All this suggests, that more than the averaged static conformation, it is the dynamical trajectories in the space of protein conformations that determine the function – and the functional trajectories seem to be constrained to low dimensional manifolds, which correspond to low-energy modes [42]. Hence, one may abandon the 'structure' as an intermediate stage, and apply a phenomenological approach in which the composed map $\pi = f \circ s$ is a single morphism from the amino acid chain to the dynamical modes. This view is in line with studies that demonstrate the functional relevance of correlations in amino acid substitutions [43-45] without the need to invoke a structural model.

The protein map $\pi = f \circ s$ starts by dimensional expansion from the linear amino acid chain to the space of conformations and their dynamical trajectories. It is followed by a dimensional reduction to a small number of relevant modes, which govern the function of the protein. An open question is whether the relevant dynamical modes are sensitive to the detailed structure of the folded protein, or whether they are coarse grained features, which are the outcome of 'low modes' in the sequence. The toy model realizes a 'protein' as a spatial network of 'amino acids' whose interaction is similar to that a neural network or a Boolean network with short-range connectivity. This put forward the notion of a network of logic operators with spatiotemporal dynamics that is encoded in genes. These metric logic networks represent the information in the DNA in terms of functional modes, which in turn operate on the DNA via numerous feedback loops. In this sense, one can think of the protein-DNA system as a molecular embodiment of the idea of self-reference.

**Acknowledgments**

The author thanks Albert Libchaber, Stanislas Leibler, and Jean-Pierre Eckmann for deep discussions and comments. This work was supported by IBS-R020-D1. TT is the Helen and Martin Chooljian Founders' Circle Member in the Simons Center for Systems Biology at the Institute for Advanced Studies, Princeton.